\documentclass[prd,showpacs,preprint,nofootinbib]{revtex4-1}
\usepackage{amsmath,amssymb}
\usepackage[colorlinks=true,linkcolor=blue]{hyperref}
\usepackage{graphicx}% Include figure files

\newcommand{\nn}{\nonumber}
\def\/{\over}
\newcommand{\bra}[1]{\langle#1|}
\newcommand{\ket}[1]{|#1\rangle}

\begin{document}

\title{Manipulating lightcone fluctuations in an analogue cosmic string}

\author{Jiawei Hu$^{1}$ and Hongwei Yu$^{1,2,}\footnote{Corresponding author at hwyu@hunnu.edu.cn}$}
\affiliation{
$^1$ Department of Physics and Synergetic Innovation Center for Quantum Effects and Applications, Hunan Normal University, Changsha, Hunan 410081, China\\
$^2$ Center for Nonlinear Science and Department of Physics, Ningbo University,  Ningbo, Zhejiang 315211, China}

\begin{abstract}

We study the flight time  fluctuations in an anisotropic medium inspired by a cosmic string with an effective  fluctuating refractive index caused by  fluctuating vacuum electric fields, which are analogous  to the lightcone fluctuations  due to fluctuating spacetime metric when gravity is quantized.  The medium can be realized as a metamaterial that  mimics a cosmic string in the sense of transformation optics. For a probe light close to the analogue string, the flight time variance is $\nu$ times that in a normal homogeneous and isotropic medium, where $\nu$ is a parameter characterizing the deficit  angle of the spacetime of a cosmic string. The parameter $\nu$, which is always greater than unity  for a real cosmic string, is determined by the dielectric properties of the metamaterial for an analogue string.
 Therefore,  the flight time fluctuations of a probe light can be manipulated % ( amplified at will)
 by changing the electric permittivity and magnetic permeability of the analogue medium.  
We argue that  it seems  possible to fabricate a metamaterial that  mimics a cosmic string with a large $\nu$ in laboratory so that a currently observable  flight time variance  might be achieved.

\end{abstract}

\maketitle

%%%%%%%%%%%%%%%%%%%%%%%%%%%%%%%%%%%%%%%%%%%%%%%%%%%%%%%%%%%%%%%%%

\section{Introduction}

Quantization of  fundamental interactions such as electromagnetic, weak and strong interactions, has achieved great success, on one hand, but on the other hand, quantization of gravity, the fundamental interaction known to the mankind the earliest, still remains elusive.
If, however, we accept  that  the basic quantum principles we are already familiar with apply as well to a quantum theory of gravity, we can make some predictions about expected quantum effects, even in the absence of a fundamental underlying theory.   The uncertainty principle is one of such principles and one generic  prediction arising from its application to the theory of gravity is that lightcones, the boundaries between spacelike and timelike regions, are no longer fixed, but smeared out due to the quantum fluctuations of spacetime metric. Since the ultraviolet divergences in quantum field theory arise from the lightcone singularities of two-point functions, it was even conjectured first by Pauli \cite{Pauli} and later investigated by other authors \cite{Deser,DeWitt,Isham71,Isham72} that the divergences might be removed when gravity is quantized. The theoretical implications and the detectability of metric fluctuations have been extensively studied \cite{Hu,Ng,Amelino-Camelia,Ellis,Borgman1,Borgman2,Polarski,Christiansen,Ford95,Ford96,Yu99,Yu00,Yu03,Yu09,Mota}.  A direct result of lightcone fluctuations is that the flight time of a probe light signal from its source to a detector spreads about the classical value in both directions~\cite{Ford95,Ford96,Yu99,Yu00,Yu03,Yu09,Mota}. These flight time fluctuations are in principle observable but are extremely small in general and seem to be undetectable in experiment in the  foreseeable future.

Therefore, it is of interest to resort to analogue systems to see whether some basic predictions of quantum gravity can be analogously verified there.  Indeed, if some such predictions can be tested, we may be able to derive useful constraints on the properties of the true underlying theory in which gravity is quantized. Based on nonlinear optics, Ford {\it et al}. proposed an analogue model for quantum lightcone fluctuations \cite{analog1,analog2}. In a nonlinear medium, the flight time of a probe light fluctuates due to a fluctuating effective refractive index  when the medium is subjected to a fluctuating background field, which is analogous to the lightcone fluctuation when gravity is quantized. The fluctuating background field can be either a single field mode in a squeezed state \cite{analog1}, or a bath of multi-mode fluctuating electromagnetic fields in vacuum \cite{analog2}. These are analogue models for active gravitational field fluctuations, which are fluctuations of the dynamical degrees of freedom of gravity itself. In Refs. \cite{analog3,analog4}, an analogue model for passive fluctuations of gravity driven by quantum stress tensor fluctuations has also been proposed.

In this paper, we study the flight time fluctuation of certain probe light pulses that arises from a fluctuating refractive index due to  electromagnetic vacuum fluctuations in an anisotropic medium that mimics a cosmic string in terms of the classical propagation of light. Such a medium can be realized as a metamaterial in experiment. In particular, we demonstrate that the flight time fluctuations can be amplified  compared with those in a normal medium by manipulating the electric permittivity and magnetic permeability of the analogue medium, and remarkably,  a currently experimentally observable  flight time variance  would be  obtained if an analogue cosmic string with a large enough $\nu$  could be fabricated in laboratory. Here let us note that the lightcone fluctuations due to metric fluctuations in the cosmic string spacetime have recently been studied in Ref. \cite{Mota}. The Lorentz-Heaviside units with $\hbar=c=1$ are used in this paper unless specified.

%%%%%%%%%%%%%%%%%%%%%%%%%%%%%%%%%%%%%%%%%%%%%%%%%%%%%%%%%%%%%%%%%

\section{The Basic Formalism}
In a nonlinear medium, the electric polarization $P_i$ can be expanded in a power series of the electric field $E_i$ as
\begin{equation}
P_i = P_i^{(1)} + P_i^{(2)}+\cdots
    = \chi_{ij}^{(1)} E^j + \chi_{ijk}^{(2)} E^j E^k + \cdots  \;.
\end{equation}
Here $\chi^{(i)}$ is the {\it i}-th order susceptibility tensor. We write the total electric field $E^i$ as a sum of a background field $E_{0}^i(\omega_0)$ and a probe field $E_{1}^i(\omega_1)$. So, the second order polarization $P_i^{(2)}$ takes the form
\begin{equation}
P_i^{(2)}(\omega_m+\omega_n)
=\sum_{m,n=0}^1\chi_{ijk}^{(2)}(\omega_m+\omega_n)E_m^j(\omega_m)E_n^k(\omega_n)\;.
\end{equation}
We assume that for the background field $E_{0}^i(\omega_0)$, the second order susceptibility tensor $\chi^{(2)}_{ijk}(2\omega_0)$ can be neglected. So, in the absence of the probe field  $E_{1}^i$, the medium the background field $E_{0}^i$ propagates in can be treated as a linear medium characterized by a dielectric tensor
$\varepsilon_{ij}$ such that $D^i=\varepsilon^{ij} E_j$  \cite{analog2}.
Here we have neglected dispersion in the frequency range of the background fields. On the other hand, the magnetization is also assumed to be linear with the applied magnetic field, so $B^i=\mu^{ij}H_j$.  In this paper, we are interested in an anisotropic  metamaterial medium in which the propagation of light rays is equivalent to that in a static, straight cosmic string spacetime, with the line element being
\begin{equation}\label{metric-cs}
ds^2=dt^2-d\rho^2-\frac{\rho^2}{\nu^{2}} d\phi^2-dz^2\;.
\end{equation}
Here $\nu=(1-4G\mu)^{-1}$, with $\mu$ the mass per unit length of the string, and $G$ the Newtonian constant of gravitation.  Let $\theta=\phi/\nu$, the line element can be rewritten as
\begin{equation}
ds^2=dt^2-d\rho^2-\rho^2 d\theta^2-dz^2\;,
\end{equation}
where $\theta \in [0,2\pi/\nu)$. This metric describes a flat spacetime with a deficit angle $8\pi G\mu$. In the framework of transformation optics \cite{trans1,trans2,trans3}, the electric permittivity and magnetic permeability tensors of a medium that mimics the propagation of light in the cosmic string spacetime take the form
\begin{equation}
\varepsilon^{ij}=\mu^{ij}
=n_B\left(
\begin{array}{ccc}
1 & 0 & 0
\\
0 & \displaystyle{\frac{\nu^2}{\rho^2}} & 0
\\
0 & 0 & 1
\end{array}
\right),
\end{equation}
in the cylindrical coordinate.  After a coordinate transformation from the cylindrical coordinate to the Cartesian coordinate, $\varepsilon^{ij}$ and $\mu^{ij}$ become
\begin{equation}\label{epsilon-nu}
\varepsilon^{ij}  = \mu^{ij} =
n_B \left(
\begin{array}{ccc}
\cos\phi^2 + \nu^2 \sin\phi^2 &
\left(1-\nu^2 \right) \cos\phi\sin\phi & ~0
\\
\left(1-\nu^2 \right) \cos\phi\sin\phi &
\sin\phi^2 + \nu^2 \cos\phi^2 & ~0
\\
0 & 0 & ~1
\end{array}
\right) \;,
\end{equation}
which is in agreement with those obtained in Ref. \cite{meta-1} for spinning cosmic strings when the angular momentum of the cosmic string approaches zero, as expected. Note that when $\nu=1$, the above tensor describes  a normal homogeneous and isotropic medium and empty space if  the refractive index of the background field  $n_B$ is further set to be unity.

The electromagnetic wave equation in such a medium can be written as
\begin{equation}\label{wave}
 -\frac{1}{\sqrt{\gamma}} \partial_j \sqrt{\gamma} \gamma^{jk} \partial_k
  \gamma^{im}E_m
 +\frac{1}{\sqrt{\gamma}} \partial_j \sqrt{\gamma} \gamma^{ik} \partial_k
   \gamma^{jm}E_m
 + \frac{1}{v_B^2}\frac{\partial^2 \gamma^{im}E_m }{\partial t^2}
=0\;,
\end{equation}
where $\gamma=\rho/\nu$ is the determinant of the spatial metric tensor of the cosmic string spacetime $\gamma_{ij}={\rm diag}(1,\rho^2/\nu^2,1)$. This equation takes the same form as that in the cosmic string spacetime with an effective speed of light $v_B=1/n_B$.

Now we consider a probe light $E_{1}^i$,  which  is much smaller than the background field $E_{0}^i$, while its frequency $\omega_1$ is much larger than that of the background field $\omega_0$ \cite{analog1,analog2}. If  the probe field is propagating in the $z$-direction and is  polarized in the $\rho$-direction, i.e. $E_{1}^i = \delta^{i\rho} E_{1}(t,z)$, the wave equation for $E_{1}$ takes the form
\begin{equation}\label{wave-p}
 -\frac{\partial^2 E_{1}}{\partial z^2}
 + \frac{1}{v_P^2} \left[1+ \frac{1}{n_P^2} \left(\chi_{\rho\rho j}^{(2)} + \chi_{\rho j\rho}^{(2)}\right) E_{0}^j \right] \frac{\partial^2 E_{1}}{\partial t^2} = 0\;,
\end{equation}
where
%\begin{equation}
%\epsilon_1 =\Gamma_j E_{0}^j \;,
%\end{equation}
%with
%\begin{equation}\label{gamma}
%\Gamma_j=\frac{1}{2n_P^2}
%\left(\chi_{\rho\rho j}^{(2)} + \chi_{\rho j\rho}^{(2)}\right) \;.
%\end{equation}
$n_P=1/v_P$.  This equation describes a wave propagating with a space and time dependent phase velocity
\begin{equation}\label{v}
v \approx v_P \left[1- \frac{1}{2n_P^2} \left(\chi_{\rho\rho j}^{(2)} + \chi_{\rho j\rho}^{(2)}\right) E_{0}^j  \right]\;,
\end{equation}
where $\left| \frac{1}{2n_P^2} \left(\chi_{\rho\rho j}^{(2)} + \chi_{\rho j\rho}^{(2)}\right) E_{0}^j \right| \ll 1$ is assumed. Generally, $v_P$ is different from $v_B$ due to dispersion. Here let us note that the unique properties of metamaterials are usually restricted to a narrow frequency range. However, in the derivation of Eq. (\ref{wave-p}), %it is not necessary to require
no assumption is made  that the medium simulates the cosmic string spacetime in the frequency regime of the probe light.

\begin{figure}[htbp]
\centering
\includegraphics[scale=0.75]{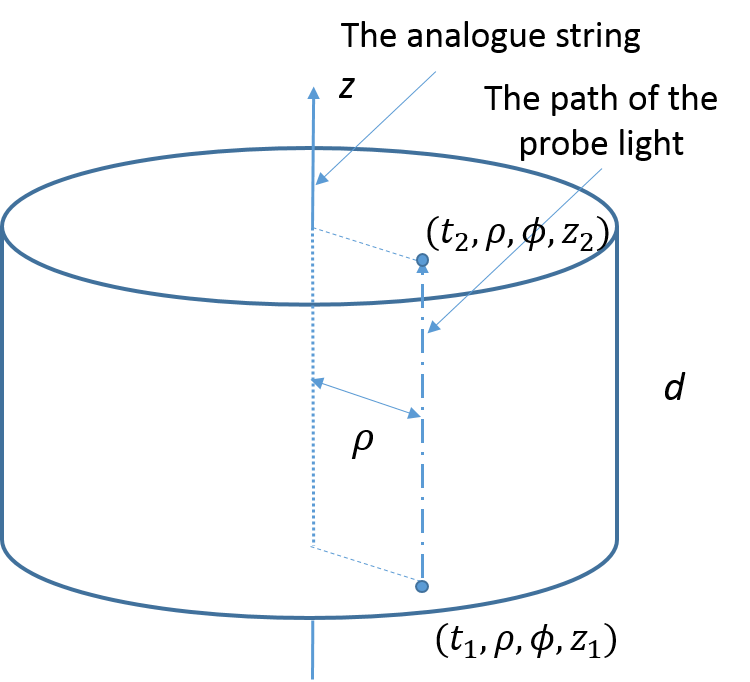}
\caption{ A probe light ray (dot-dashed line) propagates in the direction parallel to the analogue string located at the $z$-axis, from point $(t_1,\rho,\phi,z_1)$ to point $(t_2,\rho,\phi,z_2)$.}
\label{fig1}
\end{figure}

%%%%%%%%%%%%%%%%%%%%%%%%%%%%%%%%%%%%%%%%%%%%%%%%%%%%%%%%%%%%%%%%%

\section{Lightcone fluctuations in an analogue cosmic string}
Consider that the probe light propagates through the metamaterial of an analogue cosmic string located at the $z$-axis,  in  the parallel direction from $(t_1,\rho,\phi,z_1)$ to $(t_2,\rho,\phi,z_2)$, with $d=z_2-z_1$ the thickness of the material, as sketched in Fig \ref{fig1}. The flight time of the probe signal through the material can then be expressed as
\begin{equation}
t=\int_{z_1}^{z_2} \frac{dz}{v}  = n_P \int_{z_1}^{z_2} \left[1 + \frac{1}{2n_P^2}
\left(\chi_{\rho\rho j}^{(2)} + \chi_{\rho j\rho}^{(2)}\right) E_{0}^j(t,\vec{x}) \right]\, dz\;.
\end{equation}
This integration is along the path of the probe pulse, i.e. $z = v_P t= t/n_P$. In the present paper, we assume that the background field $E_0$ is the fluctuating vacuum electromagnetic field in the analogue medium of a cosmic string. The fluctuations of $E_0$ will cause the fluctuations of the flight time $t$. The relative flight time variance takes the form \cite{analog1,analog2}
\begin{equation}
\label{dt}
  \delta ^2= \frac{\langle t^2 \rangle - \langle t \rangle^2}{\langle t \rangle^2}
= \frac{1}{4 n_P^4 d^2} \int_{z_1}^{z_2} dz \int_{z_1}^{z_2} dz' \left(\chi_{\rho\rho i}^{(2)} + \chi_{\rho i\rho}^{(2)}\right) \left(\chi_{\rho\rho j}^{(2)} + \chi_{\rho j\rho}^{(2)}\right) \langle E_{0}^i(t,\vec{x})  E_{0}^j (t',\vec{x}') \rangle\;,
\end{equation}
where $\langle\,\,\rangle$ denotes the  expectation value over the vacuum state, and we have assumed that $\langle E^j_{0}\rangle=0$. Here, the vacuum expectation %$\langle\,\,\rangle$ in the equation above
is a summation over the contributions of vacuum field modes in all frequencies, %while the derivation of the effective phase velocity of the probe field Eq. (\ref{v})
while the derivation of the wave equation (\ref{wave-p})
is based on the assumption that the frequency of the background field is much smaller than that of the probe field. So, contributions from high frequency modes should be suppressed.  Fortunately, this can be realized by engineering  a smoothly varying second order susceptibility in the frequency regime of the probe light  in the direction parallel to the analogue string. To see this, let us assume that $\chi^{(2)}(z)\equiv \left(\chi^{(2)}_{\rho\rho z}(z)+\chi^{(2)}_{\rho z \rho}(z) \right)\big/2$ satisfies $\frac{1}{d}\int_{-\infty}^{\infty} dz\, \chi^{(2)}(z)= \chi^{(2)}_0$, with $\chi^{(2)}_0$ the averaged second order susceptibility along the $z$-axis~\cite{analog2}.  As an example to show that the contributions from high frequency modes are suppressed when a smooth  second order susceptibility $\chi^{(2)}(z)$ along the analogue string is assumed, we take the profile of $\chi^{(2)}(z)$ as of the Lorentzian form
\begin{equation}\label{lorentz}
\chi^{(2)}(z)= \frac{d^2}{\pi(z^2+d^2)} \chi^{(2)}_0\;.
\end{equation}
The relative flight time fluctuation (\ref{dt}) can then be reformed as
\begin{eqnarray}
 \delta^2 \propto
\bigg\langle 0\bigg |
\int_{-\infty}^{\infty} d\omega e^{-|\omega| \tau} E^i(\omega)
\int_{-\infty}^{\infty} d\omega' e^{-|\omega'| \tau} E^j(\omega')
\bigg|0\bigg\rangle\;,
\end{eqnarray}
where $E^i(\omega)$ is the Fourier transform of $E^i(t)$, and $\tau=n_P\, d$. Therefore, it is clear that contributions from vacuum modes whose wavelengths are shorter than the thickness of the medium have been suppressed. If the wavelength of the probe light is much smaller than the thickness of the medium, the assumption that the frequency of the background field is much smaller than that of the probe field can be satisfied. Note that although the background field $E_0$ is supposed to be the fluctuating electromagnetic field in vacuum, it is still possible that $E_{0}$ is much greater than $E_{1}$ \cite{analog2}.

In the following, we work out the two point function of the electric field in the anisotropic dielectric, which is needed in the calculation of the flight time variance (\ref{dt}). Since the wave equation for the background electromagnetic field is the same as that in the cosmic string spacetime, we resort to the existing result obtained in the cosmic string spacetime. Although locally the spacetime is Minkowskian, the nontrivial global  topology of the spacetime requires that the only nonzero component of the electric field on the cosmic string is that parallel to the string. This is reminiscent of an ideal reflecting boundary on which only the normal component of the electric field is nonzero \cite{zhou15}. As a result, the correlation functions that characterize the fluctuations of the electromagnetic fields are position dependent, and in the limit of small separations to the cosmic string, the only contribution to the vacuum fluctuations comes from the $z$-component of the electric field. This can be seen from Eq.~(73) in Ref. \cite{zhou15}, in which the spontaneous emission and excitation rates of an atom near the cosmic string have been studied as a probe to detect the background field fluctuations. In this work, what we are interested in is the lightcone fluctuations of a probe light whose trajectory is close to the analogue string. 
So, in what follows, we only consider the contribution from the $z$-component of the electric field $\langle E_z(t,z)E_z(t',z')\rangle$. In the cosmic string spacetime, the normal modes for the 0 and $z$ components of the vector potential in vacuum are \cite{Skarzhinsky,zhou15}
\begin{equation}
f_{0,z}(\vec{x})=\frac{1}{2\pi}\sqrt{\frac{\nu}{2\omega}}J_{|\nu m|}(k_{\bot}\rho)
e^{i(\nu m\theta+k_{3}z)-i\omega t}\;.
\end{equation}
With the help of the relation $E_i=A_{0;i}-A_{i;0}$, the electric field two point function $\langle E_z(x)E_z(x')\rangle$ can be expressed as
\begin{equation}
\langle 0|E_{z}(x)E_{z}(x')|0\rangle=
(\partial_{0}\partial_{0}'-\partial_{z}\partial_{z}')\, G^{+}(x,x')\;,
\end{equation}
where $G^{+}(x,x')$ is the scalar field two point function in the cosmic string spacetime which can be calculated as \cite{Mota}
\begin{equation}
G^{+}(x,x')= \frac{1}{4\pi^2}\frac{1}{\sigma_0^2}
            +\frac{1}{2\pi^2}\sum_{m=1}^{[\nu/2]}{'}\frac{1}{\sigma_m^2}
            -\frac{\nu}{8\pi^3}\sum_{j=+,-}\int_0^\infty d\zeta
     \frac{\sin\,[\nu(j\Delta\phi+\pi)]}{[\cosh\,(\nu\zeta)-\cos\,(j\nu\Delta\phi+\nu\pi)]}
     \frac{1}{\sigma_\zeta^2}\;,
\end{equation}
where
\begin{eqnarray}
\sigma_0^2 &=& -\Delta t^2 + \Delta z^2 + \rho^2 + \rho'^2 - 2\rho\rho'\cos\Delta\phi\;,\\
\sigma_m^2 &=& -\Delta t^2 + \Delta z^2 + \rho^2 + \rho'^2 - 2\rho\rho'\cos\left(\frac{2\pi m}{\nu}-\Delta\phi\right),\\
\sigma_\zeta^2 &=& -\Delta t^2 + \Delta z^2 + \rho^2 + \rho'^2 + 2\rho\rho'\cosh \zeta\;.
\end{eqnarray}
Here $\Delta t=t-t'-i\epsilon$, and $\Delta z=z-z'$. The $[\nu/2]$ in the summation denotes the integer part of $\nu/2$, and the prime means when $\nu$ is an even integer the $m=\nu/2$ term  should be taken with a factor $1/2$. When $\nu<2$, there is no contributions from the summation. Allowing for the fact that the net effect of a medium on the electric field two-point function is to bring an overall factor of $1/n_B^3$ and to replace the time $t$ with $t/n_B$ \cite{Glauber,Barnett,analog2}, the two-point function for the electric field in the anisotropic material that mimics a cosmic string is calculated as
\begin{eqnarray}\label{18}
 \bra{0} E_z(x)E_z(x') \ket{0}
&=&\frac{1}{\pi^2 n_B^3}
 \frac{1}{\left(\Delta t^2/n_B^2 - \Delta z^2\right)^2}+\frac{2}{\pi^2 n_B^3}\sum_{m=1}^{[\nu/2]}{'}
 \frac{\Delta t^2/n_B^2 - \Delta z^2 + 4\rho^2\sin^2\frac{m\pi}{\nu}}
 {\left(\Delta t^2/n_B^2 - \Delta z^2 - 4\rho^2\sin^2\frac{m\pi}{\nu}\right)^3}
 \nn\\
  &&-\frac{\nu}{\pi^3 n_B^3}\int_{0}^{\infty} d\zeta
 \frac{\sin \nu\pi}{\cosh \nu\zeta- \cos \nu\pi}
 \frac{\Delta t^2/n_B^2 - \Delta z^2 + 4\rho^2\cosh^2\frac{\zeta}{2}}
 {\left(\Delta t^2/n_B^2 - \Delta z^2 - 4\rho^2\cosh^2\frac{\zeta}{2}\right)^3} \;.
\end{eqnarray}
Here, the refractive index of the background field $n_B$ defines an effective lightcone $t=n_B z$. In the following, we assume that the refractive index of the probe field $n_P$ is larger than that of the background field $n_B$, so that the path of the probe light $t=n_P z$ is inside the effective lightcone%which are both inside the actual lightcone.
\footnote{ If the path of the probe light is chosen to be a ``spacelike" , i.e. $n_P<n_B$, the integral would be divergent. See Ref. \cite{Fewster15} for a discussion and more references.}.

Direct calculations with the help of the residue theorem, after inserting the two-point function (\ref{18}) into Eq. (\ref{dt}), lead to
\begin{eqnarray}\label{result-1a}
 \delta^2&=&
 \frac{ n_B \,(\chi^{(2)}_0 )^2 }{16\pi^2 (n_P^2-n_B^2)^2 n_P^4 d^4}
+\frac{ n_B \,(\chi^{(2)}_0 )^2 }{8\pi^2 (n_P^2-n_B^2)^2 n_P^4 d^4}
 {\sum_{m=1}^{[\nu/2]}}{'} \frac{1 - \tilde{\rho}^2\sin^2 \frac{m\pi}{\nu}}
 {\left(\,1 + \tilde{\rho}^2\sin^2\frac{m\pi}{\nu}\right)^3}\nn\\
&&-\frac{\nu\, n_B \,(\chi^{(2)}_0 )^2 \sin \nu\pi}{16\pi^3 (n_P^2-n_B^2)^2 n_P^4 d^4}
  \int_0^\infty d\zeta \frac{1}{\cosh \nu\zeta-\cos \nu\pi}
  \frac{1 - \tilde{\rho}^2\cosh^2 \frac{\zeta}{2}}
 {\left(\,1 + \tilde{\rho}^2\cosh^2\frac{\zeta}{2}\right)^3}\;,
\end{eqnarray}
where we have defined $\xi=\sqrt{\frac{n_P^2}{n_B^2}-1}$, and $\tilde{\rho}=\rho/(\xi d)$.  Note that when $\nu=1$, i.e.,  when the medium becomes a normal homogeneous and isotropic one, the only contribution to Eq. (\ref{result-1a}) is the first term, which is the same as that obtained in Ref. \cite{analog2}.

When the trajectory of the probe light is close to the analogue string, i.e. $\rho/d\ll 1$, the following flight time fluctuations can be derived as will be shown next
\begin{equation}\label{result-1b}
 \delta^2\approx\frac{\nu\, n_B \,(\chi^{(2)}_0 )^2 }{16\pi^2 (n_P^2-n_B^2)^2 n_P^4 d^4}\;.
\end{equation}
The derivation of the above result will be divided into two cases. First when $\nu$ is an integer, $\sin\nu\pi=0$. So, the integration in Eq. (\ref{result-1a}) does not contribute, and the derivation is straightforward by a series expansion and replacing the summation ${\sum_{m=1}^{[\nu/2]}}{'}$ with $\frac{1}{2} \sum_{m=1}^{\nu-1}$. Then, when $\nu$ is not an integer, we approximate the integration with the result when $\rho=0$, which can be calculated with the help of Eq. (3.513) in Ref.~\cite{integrate} as
\begin{eqnarray}\label{int-appr}
  \int_0^\infty d\zeta \frac{1}{\cosh \nu\zeta-\cos \nu\pi}
 =\frac{2}{\nu\, \sin\nu\pi}\arctan\cot\frac{\nu\pi}{2}\;.
\end{eqnarray}
For $\nu\geq 1/2$, it can be shown that the difference between the integration in Eq.~(\ref{result-1a}) and its approximation~(\ref{int-appr}) satisfies
\begin{eqnarray}
&&\int_0^\infty d\zeta \frac{1}{\cosh \nu\zeta-\cos \nu\pi} \,
\frac{\tilde{\rho}^6 \cosh ^6\frac{\zeta }{2}+3 \tilde{\rho}^4 \cosh^4 \frac{\zeta }{2}+4 \tilde{\rho}^2 \cosh ^2\frac{\zeta }{2}}{\left(\tilde{\rho}^2 \cosh ^2\frac{\zeta }{2}+1\right)^3}\nn\\
&&~~<
\int_0^\infty d\zeta \frac{1}{\cosh \frac{\zeta}{2}-\cos \nu\pi} \,
\frac{4\tilde{\rho}^6 \cosh ^6\frac{\zeta }{2}+8 \tilde{\rho}^4 \cosh^4 \frac{\zeta }{2}+4 \tilde{\rho}^2 \cosh ^2\frac{\zeta }{2}}{\left(\tilde{\rho}^2 \cosh ^2\frac{\zeta }{2}+1\right)^3}\nn\\
%&&~~=
%\int_0^\infty d\zeta \frac{1}{\cosh \frac{\zeta}{2}-\cos \nu\pi}
%\frac{4 \tilde{\rho}^2 \cosh ^2\frac{\zeta }{2}}{\tilde{\rho}^2 \cosh ^2\frac{\zeta }{2}+1}\nn\\
&&~~=\int_{1}^{\infty} dx
 \frac{1}{ x^2-2 x \cos \nu \pi +1 } \,
 \frac{16 \tilde{\rho}^2  \left(x^2+1\right)^2}
 {\tilde{\rho}^2 x^4+2 \left(\tilde{\rho}^2+2\right) x^2 +\tilde{\rho}^2}\;.
\end{eqnarray}
In the second line of the above, we have made the integrand larger by adding $3 \tilde{\rho}^6 \cosh ^6\frac{\zeta }{2}+5 \tilde{\rho}^4 \cosh^4 \frac{\zeta }{2}$ to the numerator and replacing $\cosh \nu\zeta$ by $\cosh\frac{\zeta }{2}$ in the denominator, and in the last line we have made a variable substitution $x=e^{\zeta/2}$. Evaluating the integration of rational functions above and series expanding the result, the leading term can be shown to be of the order of $\tilde{\rho}$.
That is, although amplified, the difference is still a higher order correction when $\rho/d\ll 1$. Therefore, for a non-integer $\nu\in(2I,2I+2)$, where $I\geq 0$ is an integer, the flight time fluctuations (\ref{result-1a}) can be evaluated as
\begin{eqnarray}
 \delta^2&\approx&
 \frac{n_B \,(\chi^{(2)}_0 )^2 }{16\pi^2 (n_P^2-n_B^2)^2 n_P^4 d^4}
 \left[1+2I-\frac{2}{\pi}\arctan\tan
 \left(\frac{\pi}{2}-\frac{\nu\pi}{2}+\pi I\right)  \right]\nn\\
 &=&\frac{\nu\,n_B \,(\chi^{(2)}_0 )^2 }{16\pi^2 (n_P^2-n_B^2)^2 n_P^4 d^4}\;.
 \end{eqnarray}
So, the flight time fluctuations are $\nu$ times those in a normal medium. %For a  real cosmic string, $\nu=(1-4G\mu)^{-1}$ is always larger than 1 since  the mass per unit length of the string $\mu>0$. However, for an analogue cosmic string, $\nu$ is determined by the dielectric properties, which can be manipulated,  so it can be either larger or smaller than 1. %Therefore,  the lightcone fluctuations can be either amplified or suppressed compared with those in a normal medium.
For a typical nonlinear medium, it has been estimated in Ref. \cite{analog2} that the root mean square of the fractional flight time fluctuation $\delta_{\rm rms}=\sqrt{\delta^2}$ is of the order of $10^{-8}$ when $d\sim 10~{\rm \mu m}$, which is potentially observable. Our result shows that the detection of the effect would become easier with an anisotropic  metamaterial that mimics a cosmic string.  In principle, an observable light cone fluctuation within the current experiment precision would be obtained if the analogue cosmic string could be fabricated with a large enough $\nu$ in laboratory.   %On the other hand, although vacuum fluctuations are inevitable due to the uncertainty principle, the lightcone fluctuations can also be suppressed in a metamaterial with  $\nu$ smaller than unity.

When the trajectory of the probe light is far from the analogue string, i.e. $\rho\gg d$, we have
\begin{eqnarray}\label{result-1c}
\delta^2&\approx&
 \frac{n_B \,(\chi^{(2)}_0 )^2 }{16\pi^2 (n_P^2-n_B^2)^2 n_P^4 d^4}\nn\\
 &&-\frac{(\chi^{(2)}_0 )^2 }{8\pi^2 n_B^3\,n_P^4\, \rho^4 }
 \left(\sum_{m=1}^{[\nu/2]}{'}\frac{1}{\sin^4\frac{m\pi}{\nu}}
     -\frac{\nu \sin\nu\pi}{2\pi}
     \int_0^{\infty}  d\zeta \frac{1}{\cosh \nu\zeta-\cos \nu\pi}
     \frac{1}{\cosh^4 \frac{\zeta}{2}} \right).
\end{eqnarray}
Therefore, when $\rho\gg d$, the corrections to the flight time variance due to the presence of an analogue cosmic string (the second term in Eq. (\ref{result-1c})) is proportional to $\rho^{-4}$, which is  a   higher-order correction compared with that in a normal  isotropic medium which is proportional to $d^{-4}$.  For an integer $\nu$, the result can be further calculated as
\begin{eqnarray}\label{result-1d}
\delta^2&\approx&
\frac{n_B \,(\chi^{(2)}_0 )^2 }{16\pi^2 (n_P^2-n_B^2)^2 n_P^4 d^4}
   -\frac{(\chi^{(2)}_0 )^2\, (\nu^4+10\nu^2-11)}{720\pi^2 n_B^3\,n_P^4\, \rho^4 }\;,
\end{eqnarray}
where we have used the relation
\begin{equation}
{\sum_{m=1}^{\nu-1}}\frac{1}{\sin^4\frac{m\pi}{\nu}}=\frac{1}{45}(\nu^4+10\nu^2-11)\;.
\end{equation}
It is interesting to note that the correction term is negative and so the flight time fluctuations of the probe light is weakened as compared with those in a normal medium.

The discussions above suggest that the flight time fluctuations for a probe light can be amplified by a factor of $\nu$ when its trajectory is close to the analogue cosmic string. Now an important question is how large the factor $\nu$ could be. Recall that $\nu$ is related to the dielectric tensor $\varepsilon$ as in Eq~(\ref{epsilon-nu}), which can be rewritten as
\begin{equation}
\varepsilon  = R^T(\phi)\, \varepsilon_U R(\phi)\;,
\end{equation}
where $R(\phi)$ is the rotation matrix around $z$ axis, 
\begin{equation}
R(\phi)=
\left(
\begin{array}{ccc}
 \cos \phi & \sin \phi & 0 \\
 -\sin \phi & \cos \phi & 0 \\
 0 & 0 & 1 \\
\end{array}
\right) \;, 
\end{equation}
$R^T(\phi)$ is the transpose of $R(\phi)$, and 
\begin{equation}\label{varepsilon-U}
\varepsilon_U =
\left(
\begin{array}{ccc}
 \varepsilon_\parallel & 0 & 0 \\
 0 & \nu^2\varepsilon_\parallel & 0 \\
 0 & 0 & \varepsilon_\parallel \\
\end{array}
\right) 
\end{equation}
is the dielectric tensor of a uniaxial dielectric. That is, such a medium can be realized with an anisotropic dielectric with a rotational symmetry around the $z$ axis, and $\nu$ describes the anisotropy of the medium. For a class of dielectrics with extreme optical anisotropy, $\nu$ can be extremely large \cite{metamaterial}. The appearance of a large $\nu$ in Eq. (\ref{varepsilon-U}) means that the uniaxial medium is perfectly conducting along the $y$ direction. Such a medium can be realized with artificial metamaterials, e.g. an array of parallel metallic wires \cite{Belov03,Silveirinha06}, or a layered-metal dielectric structure \cite{Belov06}. For an order of magnitude estimation, we assume that the wavelength of the probe light $\lambda_P$ is of the order of $\sim 1~{\rm \mu m}$, and the thickness of the metamaterial $d$ is of the order of $\sim 10~{\rm \mu m}$. Therefore, background field modes whose wavelengths $\lambda_B\gtrsim 10~{\rm \mu m}$ should be taken into account. For $\lambda_B\sim 10~{\rm \mu m}$, the refractive index for a conductor, e.g. Ag (silver), is of the order of $10$ \cite{Honghua15}, while that of a dielectric is of the order of 1, so $\nu=\sqrt{\varepsilon_{yy}/\varepsilon_{xx}}=n_{\rm Ag}/n_{\rm dielectric}\sim 10$, where we have used $n=\sqrt{\varepsilon\mu}$, and $\mu\approx1$. The ratio $\nu=n_{\rm Ag}/n_{\rm dielectric}$ may be even larger for a larger wavelength. Therefore, it seems to be possible to fabricate a metamaterial that  mimics a cosmic string with a large $\nu$ in laboratory.

%%%%%%%%%%%%%%%%%%%%%%%%%%%%%%%%%%%%%%%%%%%%%%%%%%%%%%%%%%%%%%%%%

\section{Summary}
In this paper, we have investigated the lightcone fluctuations in an anisotropic metamaterial of an analogue cosmic string. In the absence of the probe field, the medium the fluctuating background field propagates in can be treated as a linear medium that  mimics the cosmic string spacetime  in terms of the classical propagation of light. That is, in the frequency regime of the background field, the second order susceptibility can be neglected. However, when the probe field is present, the nonlinear effect has to be taken account of, which leads to flight time fluctuations due to an effective fluctuating refractive index. Moreover, the second order susceptibility for the probe light is supposed to be position dependent along the path of the probe light, which effectively suppresses the contributions from high frequency fluctuating vacuum modes as required in our model.  For a probe light  close to the analogue string, it has been shown that the flight time fluctuations of the probe light can be amplified  compared with those in a normal medium and become observable if an analogue cosmic string with a large enough $\nu$ could be fabricated in laboratory.  So, our study here may seem to offer a possible way to manipulate lightcone fluctuations,  presumably the most direct effect of quantum gravity, in analogue systems, and make it potentially observable.

{\it Acknowledgments.---} We would like to thank Hui Liu and Chong Sheng for valuable discussions. This work was supported in part by the NSFC under Grants No. 11435006,  No. 11690034,  No. 11375092, and No. 11447022;  the Zhejiang Provincial Natural Science Foundation of China under Grant No. LQ15A050001.

\end{document}